\documentclass[fleqn,10pt]{wlscirep}
\usepackage[utf8]{inputenc}
\usepackage[T1]{fontenc}

\usepackage{graphicx}%
\usepackage{eucal}
\usepackage{amsmath,amssymb,amsfonts}%
\usepackage{amsthm}%
\usepackage{mathrsfs}%
\usepackage[title]{appendix}%
\usepackage{xcolor}%
\usepackage{textcomp}%
\usepackage{manyfoot}%
\usepackage{booktabs}%
\usepackage{algorithm}%
\usepackage{algorithmicx}%
\usepackage{algpseudocode}%
\usepackage{listings}%
\usepackage{graphicx}
\usepackage{caption}
\usepackage{setspace}
\usepackage{epstopdf}
\usepackage{makecell}
\usepackage{multirow}
\usepackage{multicol}
\usepackage[normalem]{ulem}
\usepackage[version=4]{mhchem}
\usepackage{textcomp}

\usepackage[backend=biber,style=nature,autocite = superscript,maxnames=99]{biblatex}
\usepackage{lineno}
\newtheorem*{proposition*}{Proposition}%
\newtheorem*{proof*}{Proof}

\makeatletter
\renewcommand{\fnum@figure}{Fig. \thefigure \space|}

\newif\ifsubmit
\submitfalse

\ifsubmit
    \newcommand{\can}[1]{}
    \newcommand{\todo}[1]{}
    \newcommand{\tocite}[1]{}
    
\else
    \definecolor{comments}{rgb}{0.1, 0.66, 0.1}
    \newcommand{\can}[1]{[{\color{comments}CL: #1}]}
    \newcommand{\todo}[1]{[{\color{red}TODO: #1}]}
    \newcommand{\tocite}[1]{[{\color{red}citation-#1}]}
\fi

\addtocontents{toc}{\protect\setcounter{tocdepth}{-1}}
\begin{document}


\title{
Fault-Free Analog Computing with Imperfect Hardware
}




\author[1]{Zhicheng Xu}
\author[1]{Jiawei Liu}
\author[1,2]{Sitao Huang}
\author[1]{Zefan Li}
\author[1]{Shengbo Wang}
\author[1]{Bo Wen}
\author[1]{Ruibin Mao}
\author[1]{Mingrui Jiang}
\author[3]{Giacomo Pedretti}
\author[3]{Jim Ignowski}
\author[1]{Kaibin Huang}
\author[1,4*]{Can Li}

\affil[1]{Department of Electrical and Electronic Engineering, The University of Hong Kong, Hong Kong SAR, China}
\affil[2]{Department of Engineering Science, University of Oxford, Parks Road, Oxford, Oxfordshire, OX1 3PJ, United Kingdom}
\affil[3]{Hewlett Packard Labs, Hewlett Packard Enterprise, Milpitas, CA, USA}
\affil[4]{Center for Advanced Semiconductor and Integrated Circuits, The University of Hong Kong, Hong Kong SAR, China}
\affil[*]{E-mail: canl@hku.hk}

\fancyhf{}
\fancyfoot[R]{\thepage/\pageref{lastpage-main}}
\begin{refsection}



\begin{abstract}

The surging demand for computational power, particularly for edge computing and AI, drives research into alternative paradigms like analog in-memory computing using memristors.
These approaches compute with physical laws and overcome data movement bottlenecks by performing computations directly within memory.
However, the inherent susceptibility of analog systems to device failures (stuck-at-faults) and variations critically limits their precision and reliability. 
Existing fault-tolerance techniques, including redundancy and retraining, often prove insufficient or impractical for many high-precision applications, general-purpose computations involving fixed, non-trainable matrices, and AI scenarios where privacy is a concern.
Here, we introduce and experimentally demonstrate a fault-free matrix representation where any target matrix is decomposed into a product of two adjustable sub-matrices programmed onto the analog hardware. 
This indirect, adaptive representation allows mathematical optimization to bypass faulty devices and eliminate differential pairs, significantly enhancing computational density. 
Our memristor-based system achieved over 99.999\% cosine similarity for a Discrete Fourier Transform matrix despite a 39\% device fault rate - a fidelity that conventional direct representation cannot achieve, failing with only one single device (a 0.01\% rate). 
Furthermore, we demonstrated a 56-fold bit-error-rate reduction in a wireless communication prototype and over 196\% density and 179\% energy efficiency improvements in cross-domain benchmarks compared to state-of-the-art techniques. 
This method, validated on memristors, is broadly applicable to other emerging memories and non-electrical, photonic, and quantum computing substrates. 
Thus, this work shows that the device yield and inherent imperfections are no longer the primary critical bottlenecks in emerging analog computing hardware, enables highly robust and efficient signal processing, communication and AI on the network edge.

\end{abstract}



\maketitle

\section*{Introduction}

The explosive growth in computational demand, driven by the need to process massive sensor signals and images on the network edge\autocite{dang2024reconfigurable}, power high-speed communication networks\autocite{zeng2023realizing}, and drive ever-larger generative artificial intelligence models\autocite{buchel2025efficient}, is pushing the boundaries of current computing capabilities. 
This escalating need has driven intensive research into alternative computing paradigms that harness diverse physical phenomena directly for computation. 
Promising approaches include leveraging electrical laws within analog memory devices like memristors for in-memory computing, manipulating light according to optical principles for optical computing, and exploiting quantum mechanics for quantum computing. 
Most demonstrations focus on matrix operations, because these operations forms the computational cores of image processing\autocite{zhao2023energy,li2018analogue,zhao2021implementation}, communication\autocite{zeng2023realizing,mannocci2025analog}, scientific computing\autocite{song2024programming,zidan2018general,li2023sparse} and machine learning tasks\autocite{pei2019towards,le202364,ambrogio2023analog,hu2018memristor,wang2018fully,li2018efficient,aguirre2024hardware,ye202328,wan2022compute,khwa2025mixed,chen2020wafer,wang2022memristive}. 
Furthermore, analog computing with physics directly in the memory also fundamentally eliminates the costly data shuttling inherent in von Neumann systems, offering pathways to potentially overcome the limitations of conventional digital systems\autocite{kendall2022scalable,le202364,khwa2025mixed,ambrogio2023analog,wang2025dual,zheng2024improving,ling2024rram}.

However, despite their potential, a common fundamental challenge confronts many of these emerging technologies: inherent errors and imprecision in computation.
Unlike mature digital systems that operate with high fidelity, computations directly exploiting different physical laws are often vulnerable to defects intrinsic to the underlying devices or processes, noise, and variations.
For analog computing systems with memristors for matrix operations\autocite{strukov2008missing,yang2013memristive}, each matrix element is represented by the conductance of a memristor device, or the conductance difference of a differential pair to represent both positive and negative values with non-negative physical properties (conductance in this case).
Here, the potential for device failures or stuck states, along with unavoidable variations in device fabrication and material properties, can lead to significant deviations from the intended computation\autocite{kim2025efficient}.
While certain applications, such as pattern recognition tasks using simple networks, can tolerate some level of inaccuracy, modern models with deeper structures and more complex architectures - including large language models (LLMs), high-fidelity image processing, and communication systems - are highly sensitive to these errors.
To address these inaccuracies, recent advances propose algorithm-circuit-device co-design that can be categorized into two categories: redundancy\autocite{xia2017stuck,pedretti2021redundancy,yousuf2025layer} and retraining\autocite{li2018efficient,mao2022experimentally,liu2017rescuing,chen2017accelerator,gaol2021reliable,alibart2013pattern}. 
Redundancy used additional devices to average out errors\autocite{xia2017stuck,yousuf2025layer}, compensate for inaccuracies\autocite{song2024programming,pedretti2021redundancy}, or encode information to detect signficant computing outliers\autocite{roth2020analog,li2020analog}, yet its efficacy scales sublinearly with pre-allocated overhead (e.g. in state-of-the-art approaches, 6x redundancy still results in a 5\% accuracy drop in fault-tolerant neural network tasks). 
Retraining adjusts weights to compensate for device non-idealities, either directly in hardware\autocite{li2018efficient,yao2020fully,alibart2013pattern}, or through hardware-aware models\autocite{mao2022experimentally,liu2017rescuing,chen2017accelerator,gaol2021reliable}, but is restricted to neural networks due to tunability requirements. 
Both methods struggle in more general-purpose applications.
For example, high-precision scientific computing relying on fixed, pre-defined matrices— such as the discrete Fourier transform matrix for signal processing, finite difference matrices for partial differential equations\autocite{zidan2018general}, and matrices in Krylov-subspace methods for solving equations \autocite{le2018mixed} — cannot adapt to co-design approaches because their weights must remain unchanged.
Furthermore, the hardware-aware re-training of neural networks is often impractical due to privacy constraints with training data and the substantial computational resources required by large-scale models such as LLMs\autocite{xia2024understanding}.

In this work, we propose and implement a new matrix representation tolerant to faulty physical substrates, like memristors, by representing any target matrix as a product of two adjustable sub-matrices. 
This indirect representation makes each element a collective computing from multiple devices, allowing optimization to mathematically bypass stuck devices, and eliminate the need for differential pairs to increase the density.
Using this method, representing a discrete Fourier Transform (DFT) matrix, even with a 39\% stuck-at-OFF rate, the system maintains near-perfect matrix representation, with a cosine similarity >99.999\%, a fidelity that conventional direct matrix representation method fails to achieve with even one single stuck device (0.01\% stuck-at-OFF rate).
By imposing constraints during this optimization, we also eliminate differential pairs, significantly enhancing computational density and energy efficiency. 
In image processing demonstrations using 2D-DFT, our method enhanced reconstruction quality by 15.9 dB ($39\times$ in signal-to-noise ratio) while simultaneously reducing the device count by 46.9\% compared to traditional methods. 
Furthermore, in a prototype wireless communication system, we drastically reduced the bit error rate (BER) on the faulty chip from 5.227\% to just 0.094\%, a 56-fold reliability improvement that approaches the digital baseline (0.047\%).  
Cross-domain benchmarking (including image processing, PageRank, and AI inference) consistently revealed density improvements exceeding 196\% and energy savings over 179\% compared to state-of-the-art approaches, all while effectively mitigating faulty devices. 
Although we demonstrate this method using memristor hardware as proof of concept, it can be readily applied to other technologies, including various emerging memories (phase change\autocite{burr2015experimental,le202364,joshi2020accurate,ambrogio2023analog}, ferroelectric\autocite{li2024high,soliman2023first}, spintronic\autocite{chiu2023cmos,you202514}) and non-electrical computing substrates such as photonic\autocite{zhou2022photonic} and quantum systems. 
This work thus presents a validated, practical pathway towards robust and efficient analog in-memory computing on imperfect hardware.

\section*{Fault-Free Matrix Represented by Faulty Crossbars}
\label{sec:MainIdea}
\begin{figure}[H]
\centering
\includegraphics[width=0.95\textwidth]{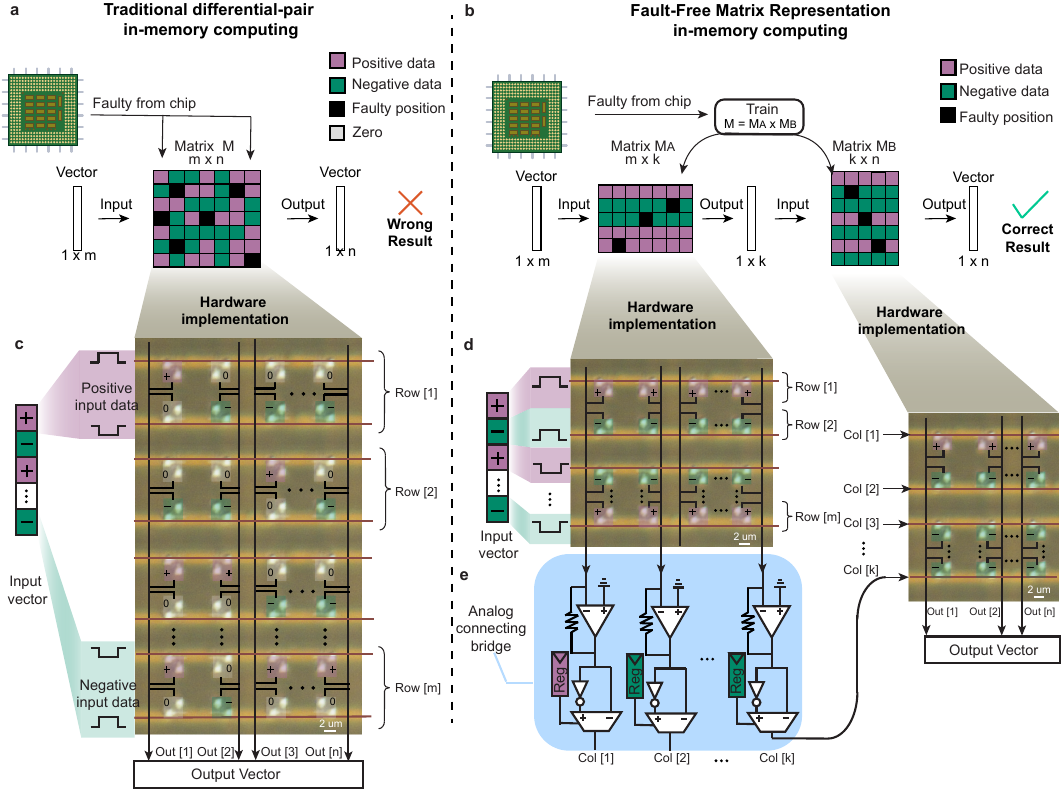}
\caption{
{\textbf{Overview of Fault-Free Matrix Representation}} 
{\textbf{a},} The traditional non-adaptive differential-pair based program. When encountering faulty devices, its output deviates from the correct one. {\textbf{b},} The basic principles of the proposed fault-free matrix representation, where any matrix is expressed as the product of two matrices, with values matching the conductance of faulty devices. 
Additional constraints can be chosen to enforce that matrix element values in the same row have the same sign to eliminate differential-pair. 
{\textbf{c},} The hardware implementation on a memristor-based chip employs differential device pairs to represent matrix values, enabling the representation of both positive and negative values with non-negative conductance. Positive input voltage is applied on top, negative input voltage below, and the result is their current summation. 
{\textbf{d},} The proposed matrix representation with adaptively decomposing into two submatrices to tackle the impact of stuck-at-fault devices.  
Two crossbars are connected by an analog bridge, and results are converted to digital signals at the final output, causing minimum additional overhead. 
The inputs voltage polarity is based on the sign corresponding to the all-positive/negative row in the decomposed matrices. 
In the case of vector input and matrix row both being positive/negative, the product output should be positive; in the case where the signs of vector and matrix row are different, their product output should be negative. 
{\textbf{e},} Detail of proposed analog-connecting bridge design. 
Register selects the sign of the output to correspond to the sign of the row from the lower decomposed matrix. The sign is determined from the decomposition process. It controls the multiplexer such that the analog input to the lower decomposed matrix is appropriate.}

\label{Fig:MainIdea}
\end{figure}

\autoref{Fig:MainIdea}a shows a memristor-based crossbar as an in-memory analog computing hardware to accelerate the processing of vector-matrix multiplication, which is the computational cores of signal / image processing, next generation communication, and AI. 
Traditionally, the weight matrix for computing is represented by the conductance of the analog memory devices at the crosspoint. 
However, some devices may be unresponsive to programming pulses and become stuck at a certain conductance value, either low (stuck at OFF) or high (stuck at ON), which can significantly degrade the accuracy of in-memory computing operations, thereby limiting its scope of application (\autoref{Fig:MainIdea}a). 
This issue affects all analog-grade memories due to aging and is more prominent in emerging technologies because of their intrinsic stochasticity, potential fabrication errors, or other imaturities\autocite{huangfu2017computation,xia2017stuck,chen2014rram}. 
Our idea is based on the observation that one matrix can be decomposed into and thus represented by a product of another two matrices, i.e., $\mathbf{M}_{m \times n} = \mathbf{M_A}_{m \times k} \mathbf{M_B}_{k \times n}$, illustrated in \autoref{Fig:MainIdea}b ,and in this way, the vector matrix multiplication can be calculated by two sub-matrix multiplications, or $\mathbf{y}=\mathbf{x}\mathbf{M}_{m \times n} = \left( \mathbf{x}\mathbf{M_A}_{m \times k} \right) \mathbf{M_B}_{k \times n}$.
The choice of the two sub-matrices is not unique and therefore can be adjusted under certain constraints, such as to match the conductance of the faulty devices.

The constraints can be expanded to allow for additional functionality, in addition to handling faulty devices. 
For example, we can eliminate the use of differential pairs in the analog crossbar, a common practice in analog computing to represent both positive and negative values with non-negative device conductances. 
In \autoref{Fig:MainIdea}c, using differential pair programming, two rows of devices represent a single row value in the matrix: one for positive data and another for negative. The input voltage signals applied to these two adjacent rows have identical amplitudes but opposite polarities. 
Mathematically, this method is expressed as: $I_{j,out} = \sum_i \left(V_i G^+_{i,j} + (-V_i) G^-_{i,j}\right) =\sum_i V_i(G^+_{i,j}-G^-_{i,j})$. 
Despite its simplicity, this approach requires twice the number of devices to represent a single value in the matrix, which can be a significant overhead for large matrices.
By decomposing the matrix into two sub-matrices, we can enforce that matrix element values in the same row have the same sign (\autoref{Fig:MainIdea}d), and apply the opposite polarity of input for multiplying with negative values. 
Mathematical proof demonstrates that at least one solution exists for the decomposition of any matrix $\mathbf{M}_{m \times n}$ under the given constraints (see Supplementary Note 3.4).
Clearly, this method can reduce the number of devices required to represent a matrix by half, increasing the memory density and energy per task of the in-memory computing system.

For details of implementation, we use stochastic gradient descent in software to optimize the two decomposed sub-matrices under the specified constraints. 
These constraints can be adjusted and combined to meet design requirements. In other words, the elimination of faulty device effects and the differential-pair can be simultaneously achieved.
Notably, this process is fundamentally a mathematical optimization of matrix representations. As a result, only the specific constraints and matrix weights are required, eliminating the need for pre-collected datasets when programming neural networks.
More details of our optimization method are provided in Methods. 

Despite eliminating the faulty device and offering higher memory and computation density, this method requires two matrix multiplication steps in series. 
It's important to note that the two crossbars representing the sub-matrices can be connected without analog-to-digital (A/D) conversion, minimizing the additional latency overhead since A/D conversion is the major consumer \autocite{yao2020fully}. 
\autoref{Fig:MainIdea}e illustrates the circuit design for one analog computing unit, which includes two sub-crossbars of size \( m \times k \) and \( k \times n \). 
An analog buffer connects the two crossbars, converting the output current of the first crossbar to the input voltage of the second. 
A multiplexer with a 1-bit register structure selects the sign of the row in the second crossbar. If the row of the second crossbar is positive, the register data is 1, and the circuit selects the '+' output; otherwise, the '$-$' output, which is the negative of the '+' output voltage, is chosen.
For a matrix larger than the size of a single computing tile, it will be divided and programmed into several subtiles 
, and the results are processed in either the analog or digital domain.

\newpage
\section*{Performance Analysis of Fault-Free Matrix Representation}
\label{sec:analysis}
\begin{figure}[!h]
\centering
\includegraphics[width=0.95\textwidth]{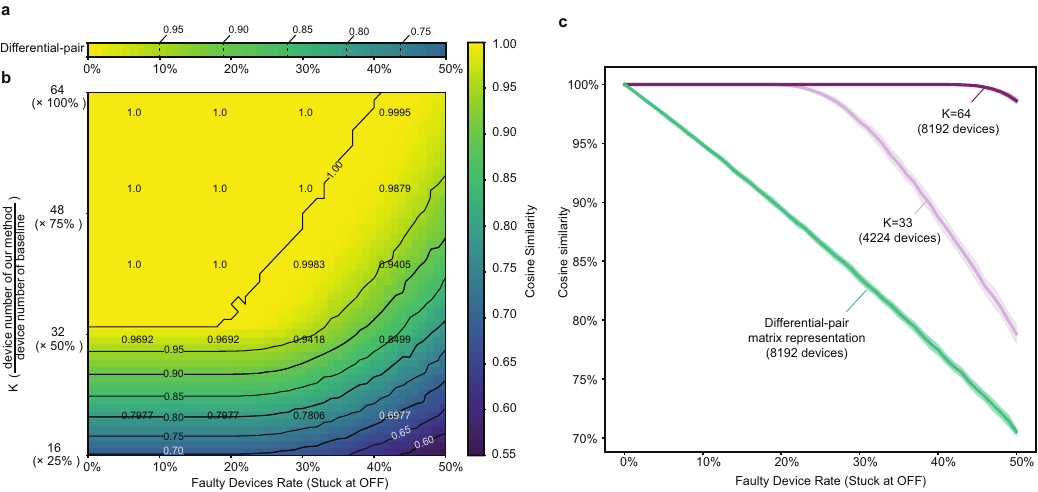}

\caption{
{\textbf{Matrix Representation Quality under Different Stuck-at Fault Ratios and Representation Configurations}} 
{\textbf{a},} and {\textbf{b},} cosine similarity heatmaps, which shows the quality of the matrix representation (more closer to 1, the better), decreases with the ratio of stuck-at-OFF devices, for both (a) conventional non-adaptive differential pair representation, and (b) our proposed adaptive matrix representation. 
In \textbf{b}, the Y-axis denotes $K$, with device usage for our method listed as a percentage of that required by the differential-pair Matrix Representation. 
Our method achieves > 99.999\% similarity at an 18\% fault rate using only 4,224 devices, significantly outperforming the differential-pair Matrix Representation, which reaches only 90\% similarity even with 8,192 devices. 
{\textbf{c},} Monte Carlo simulations with 50 trials of random stuck-at-fault device locations, but different rate and selected matrix representation configurations, including the conventional non-adaptive differential pair representation, and adaptive representation with $k=33$ and $k=64$ respectively.
When fault rates are below 18\%, both K=33 and K=64 maintain accuracy with consistent minimal degradation (<0.001\%).
}
\label{Fig:Analysis}
\end{figure}

A central consideration in our matrix representation method is the choice of the hyperparameter $k$, which defines the size of the two factorized matrices—namely, the column dimension of the first decomposed matrix and the row dimension of the second. 
A larger $k$ provides greater flexibility in adjusting weight values in the two decomposed matrices, improving robustness against a higher number of faulty devices.
Conversely, a smaller $k$ leads to using fewer devices and therefore reduces resource overhead, resulting in a more area- and energy-efficient system. 
This creates an important trade-off when selecting $k$: too small a value compromises fault tolerance, while an unnecessarily large value wastes system resources.

To validate this trade-off, we mapped the real part of a 64-point Discrete Fourier Transform (DFT) matrix to a memristor crossbar with varying stuck-at-fault rates and $k$ values (see Methods), using our proposed method.
The DFT matrix is widely used in scientific computing and has pre-defined coefficients that cannot be adjusted to compensate for device faults, making it an ideal candidate to demonstrate our method's fault tolerance capabilities.
\autoref{Fig:Analysis}b illustrates how matrix representation fidelity (measured by cosine similarity between the represented matrix and the original DFT matrix, where closer to one is better) varies with different $k$ values and ratios of stuck-at-off devices.
We also performed Monte Carlo simulations (\autoref{Fig:Analysis}c) with 50 random stuck-at-OFF fault locations at each fault rate to establish statistical validity.
From the result, one sees that, as expected, the representation fidelity decreases with increasing stuck-at-OFF fault rates, for both our methods and the conventional non-adaptive differential-pair based direct representation (baseline).
When using the same device count as the baseline (8192 devices with $k$=64), our method shows significantly improved fault tolerance—achieving almost perfect matrix representation (cosine similarity >99.999\% or cosine difference <0.1‱) even at a stuck-at-OFF fault rate of 39\%.
As a comparison, the baseline's representation error is at 1.3‱ with only one stuck-at-OFF fault device in the entire array (equally to a mere 0.01\% rate)
At the same stuck-at-fault rate of 39\%, the baseline's representation error increases to 2196‱, which is not acceptable for most applications. 
Our method also introduces $k$ as an additional tuning parameter to balance fault tolerance with resource requirements.
Reducing $k$ from 64 to 33 decreases the device count to 4224 (51.6\% of the baseline) while still maintaining high fault tolerance, with a maximum stuck-at-OFF fault rate of 18\% at 99.999\% cosine similarity.
Similar performance trends were observed with stuck-at-ON faults (Figure S2).
These simulation results demonstrate that our method achieves high fault tolerance with fewer devices, and that $k$ can be tuned to optimize the trade-off between fault tolerance and resource consumption (chip area, energy, etc.). 

\section*{Fault-tolerant and Precision-controlled Analog Computing Experiments}

\begin{figure}[H]
\centering
\includegraphics[width=0.95\textwidth]{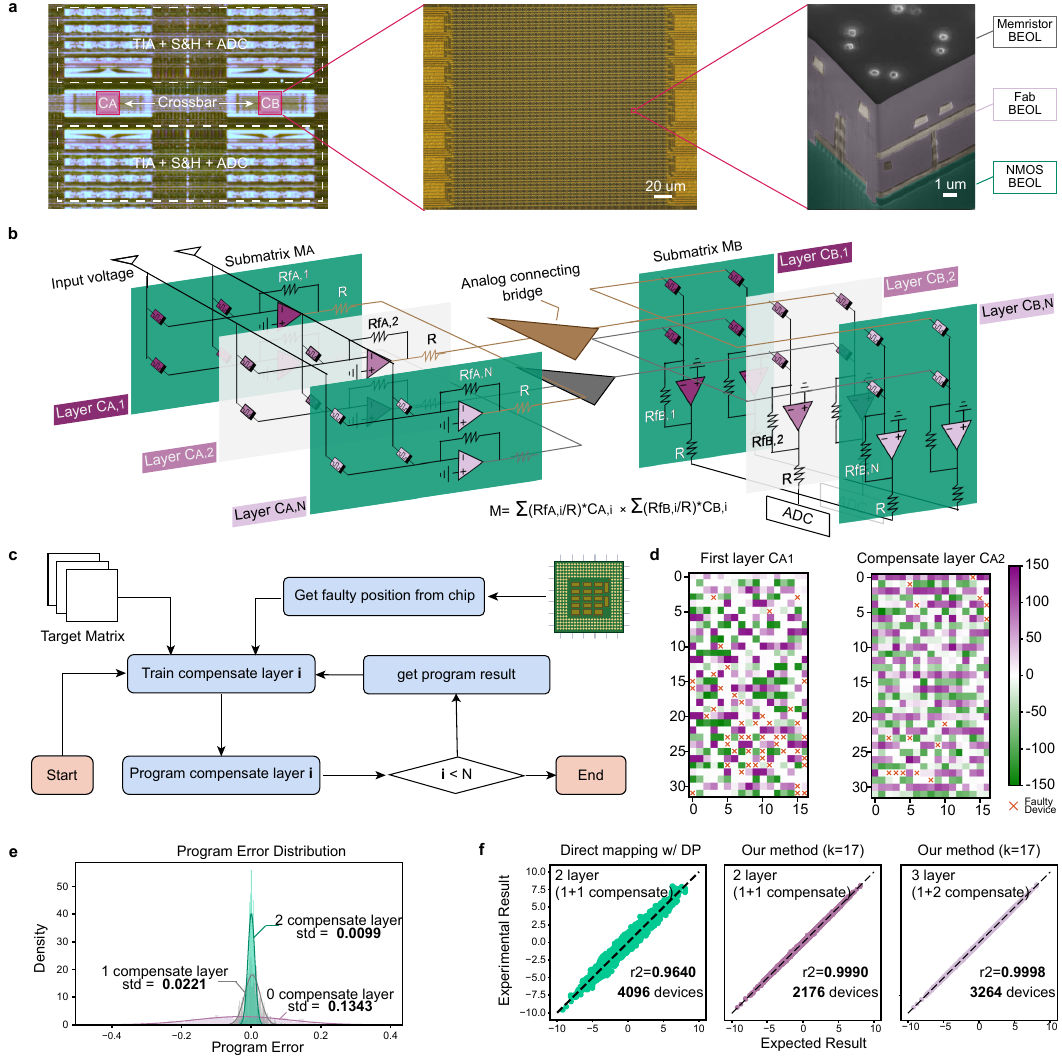}
\caption{{\textbf{Fault-tolerant and Precision-controlled Analog Computing Experiments}} 
\textbf{a,} Photograph of the integrated memristor chip, comprising multiple 64 × 64 one-transistor-one-memristor (1T1M) arrays (two crossbar marked as $C_A$ and $C_B$ are used to adaptively represent one matrix).
The photo also shows the integrated peripheral circuits, including transimpedance amplifiers (TIAs) and analog-to-digital converters (ADCs). 
\textbf{b,} Proposed hardware architecture for fault-tolerant and also precision-controlled matrix representation, where submatrices $\mathbf{M_A}$ and $\mathbf{M_B}$ are implemented using multiple compensation layers and interconnected via an analog bridge. 
\textbf{c,} Flowchart illustrating the fault-tolerant and precision-controlled programming process. 
\textbf{d,} Example of programming the real part of a 32-point DFT using a compensation layer (depicting only submatrix $\mathbf{M_A}$). 
\textbf{e,} Programming error distributions for a 32-point DFT with 0, 1, and 2 compensation layers, yielding standard deviations of 0.1343, 0.0221, and 0.0099, respectively. 
\textbf{f,} The expected and experimental vector-matrix multiplication using random inputs with different matrix representation configurations: direct mapping with differential pairs (DP) with one compensation layer, and proposed method (k=17) with one and two compensation layers, respectively. 
The results show that only the compensation layer is insufficient to overcome the significant outliers from the stuck-at-fault devices, while the proposed method effectively manage these outliers, achieving higher computing accuracy with a reduced number of devices.
}

\label{Fig:FFMR_arbitrary}
\end{figure}

The idea is experimentally demonstrated in our integrated memristor chip containing multiple tiled 64×64 crossbar arrays\autocite{li2020cmos}, with two decomposed submatrices programmed on different arrays on the same chip, respectively (shown in \autoref{Fig:FFMR_arbitrary}a).
The 50 nm $\times$ 50 nm Ta/\ce{TaO_x}/Pt memristor devices were fabricated in-house via back-end-of-line (BEOL) processing on top of 180 nm CMOS transistors, which include both selectors and peripheral circuits.
The normal working range for these memristor devices spans from 0 to 150 $\mu S$.
However, due to device aging, approximately 5\% of the devices in this chip are in a stuck-at-OFF state (unresponsive to programming pulses and permanently at 0 $\mu S$).
We deliberately selected this chip because its extensive usage history—having been powered on continuously for four years—resulted in a significant number of stuck devices, providing a challenging yet ideal scenario to demonstrate the efficacy of our adaptive matrix representation method in compensating for such hardware imperfections.
The fabrication details are documented in Methods, and the electrical characteristics are presented in Figure S3.

While the proposed fault-free matrix representation using adaptive matrix decomposition effectively eliminates the impact of faulty devices and reduces resource requirements, experimental computing precision remains limited by other non-idealities, particularly inaccurate programming of device conductance due to device-to-device and cycle-to-cycle variations.
Therefore, in our experiment, we incorporate our recent method \autocite{song2024programming} to address this inaccurate programming (not stuck devices).
This method uses compensation matrices with increasing scaling factors to correct deviations in programmed values, achieving effectively arbitrarily high programming accuracy for the conductance matrix. 
However, that method encounters difficulties with stuck devices because of the limited scalability of the compensation matrix.
Our new method presented here is fully compatible with this compensation technique. It therefore benefits from both approaches, enabling analog computing that is fault-tolerant and uses arbitrarily high-precision matrices.

To achieve both fault-tolerant and precision-controlled programming, we expanded the matrix representation $\mathbf{M}_{m\times n}
=\mathbf{M_A}_{m\times k}\times \mathbf{M_B}_{k\times n} $ to $\mathbf{M}=\sum_i^N{{k_A}_i \mathbf{C_A}_{i,m\times k}}\times \sum_i^N{{k_B}_i \mathbf{C_B}_{i,k\times n}} $, where each submatrix  $\mathbf{M_A}$ and $ \mathbf{M_B}$ is implemented through multiple parallel compensation layers ($\mathbf{C_{A}}_{1,m\times k}$--$\mathbf{C_{A}}_{N,m\times k}$ and $\mathbf{C_{B}}_{1,k\times n}$--$\mathbf{C_{B}}_{N,k\times n}$) and corresponding scaling ratio $k$. This formulation enables adaptive decomposition to mitigate faulty devices, while compensation layers correct conductance deviations, ensuring high precision.
\autoref{Fig:FFMR_arbitrary}b shows our proposed hardware architecture to implement this formulation.
In this architecture, both the current summation across compensation layers and the connection between the two adaptively decomposed parts are implemented using analog circuits, avoiding expensive analog-to-digital conversion. A detailed description of the implementation can be found in Methods. 
To configure the memristor devices in this architecture to address the faulty device issue, more constraints need to be considered. 
First, the configurations of layers $\mathbf{C_A}_{i,m\times k}$ and $\mathbf{C_B}_{i,k \times n}$ are adaptively adjusted based on the distribution of faulty devices in the current layer (\autoref{Fig:FFMR_arbitrary}c, and also the programming result of the previous layer, further detailed in Methods). 
Second, the row signs in subsequent layers should be aligned with those in the previous layer, enabling shared input voltages across layers. 
There is an alternative design choice, where the compensation is applied directly to the whole matrix, such that $\mathbf{M}_{m\times n}=\sum_i^N k_i (\mathbf{M_A}_{i,m\times k} \times \mathbf{M_B}_{i,k\times n} )$. 
However, we found that this method cannot be effectively extended to arbitrary high precision when $k < n$. 
Although we can find solutions for $\mathbf{M}_{m\times n}=\mathbf{M_A}_{m\times k} \times \mathbf{M_B}_{k\times n}$, the error compensation equation $\mathbf{M}-\mathbf{M}_{programmed}=\mathbf{M_A}_{i,{m\times k}} \times \mathbf{M_B}_{i,{k\times n}}$ may not always have a solution (mathematically, a solution must exist if $k > n$).

The effectiveness of our combined method in addressing both stuck devices and variations is demonstrated experimentally on the integrated chip, as shown in \autoref{Fig:FFMR_arbitrary}d.
We programmed the real part of a 32-point discrete Fourier transform (a $32 \times 32$ matrix) using various representation methods. 
\autoref{Fig:FFMR_arbitrary}d shows the experimentally programmed conductance values for $C_{A1}$ and the corresponding compensation $C_{A2}$ with our proposed adaptive matrix representation with interconnection rank $k=17$, as an example.
It is evident that there is a significant amount of stuck-at-OFF devices (marked `x'); the adaptive decomposition constrains these devices to remain at zero conductance.
Note also that the sign of the represented value within the corresponding row is consistent, so that to share the same voltage input across different compensation layers. 

Due to the stuck devices, conventional direct matrix mapping results in significant computing errors (Figure S6a). 
While incorporating compensation layers (analog slicing \autocite{song2024programming}) improves accuracy, substantial computing outliers persist, dominated by the high ratio of stuck devices on our chip (Figure S6b). 
Our adaptive decomposition method effectively eliminates these significant outliers, leading to much-improved computing accuracy even with fewer devices (Figure S6c, e) and a near-Gaussian error distribution (Figure S5b, d).
However, accuracy is still limited by residual programming inaccuracies in the non-stuck devices. 
Comparing Figure S6c and Figure S6e reveals that a smaller intermediate dimension $k$ (meaning matrix represented using fewer devices) makes the computation more susceptible to these programming inaccuracies.
Combining both adaptive decomposition and compensation layers tackles both faulty devices and programming inaccuracy, as demonstrated in \autoref{Fig:FFMR_arbitrary}e and Figure S5. 
The near-Gaussian error distribution confirms the elimination of outliers, while additional compensation layers offer finer control over precision, with more compensation layers resulting in higher precision.
The comparison in \autoref{Fig:FFMR_arbitrary}f demonstrate the superiority of our proposed method. 
Even when using fewer devices, our adaptive representation with compensation layers significantly outperforms conventional direct mapping, even when the latter also uses compensation. 
We achieved \(r^2\) values of 0.9990 (adaptive, k=17, one compensation layer, 2,176 devices) and 0.9998 (adaptive, k=17, two compensation layers, 3,264 devices), compared to 0.9640 for direct mapping with differential pairs and one compensation layer (4,096 devices).
Adding more compensation layers can further enhance results by mitigating more device variation, though the benefits do not scale proportionally to the increase with the increase in chip area, energy consumption, etc.

\section*{High-fidelity Image Processing Experiments}

\label{sec:experiment_image}
\begin{figure}[H]
\centering
\includegraphics[width=\textwidth]{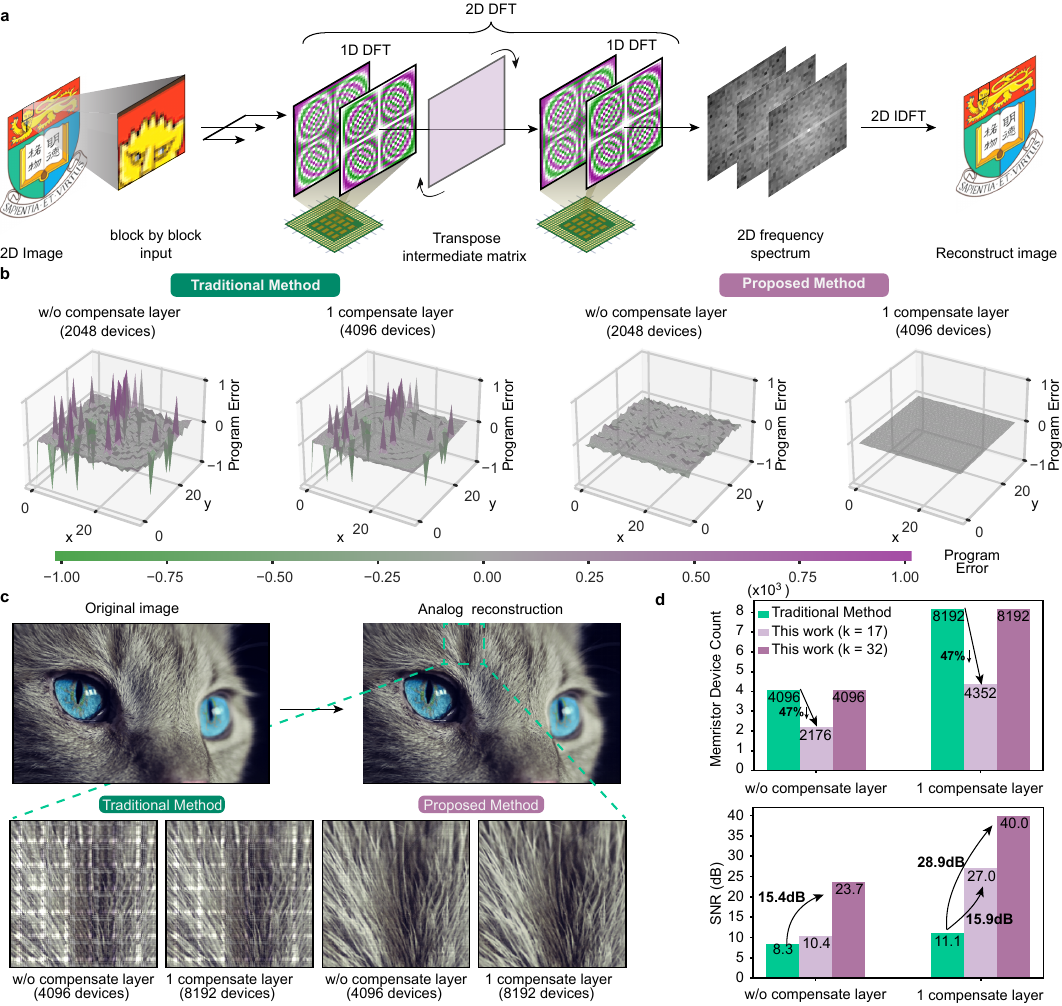}
\caption{{\textbf{Image Processing with Memristor Chip with Faulty Devices}} 
{\textbf{a},} Schematic of the in-memory 2D-DFT operations for image processing, showing how block-based processing, with sequential 1D-DFT operations along rows and columns, and a transpose step, to accommodate crossbar size constraints. 
{\textbf{b},} 
Error distributions for the real part of the memristor represented DFT matrix with different methods, showing significantly reduced errors with our design (right) compared to the traditional method (left).  
{\textbf{c},} Reconstructed images using the traditional and proposed methods with and without compensation layers, demonstrating sharper details and fewer artifacts with our design. 
Both methods utilized the same memristor device count to represent the matrix for a fair comparison.  
{\textbf{d},} 
Quantitative analysis of FFMR improvements: 
When device usage decreases by 47\%, our design improves the signal-to-noise ratio (SNR) by 15.9 dB. Moreover, when maintaining the same number of devices as the traditional method, our design achieves a superior SNR improvement of 28.9 dB.
The $k = 17$ configuration represents the decomposition of both the real and imaginary parts into the product of two matrices sized 32×17 and 17×32, while the $k = 32$ configuration represents their decomposition into the product of two 32×32 matrices.
}
\label{Fig:app1}
\end{figure}
Having experimentally validated our method on the platform, we now apply it to real-world applications. 
The first application to demonstrate is image processing using two-dimensional Discrete Fourier Transform (2D-DFT). 
The 2D-DFT is a widely utilized technique for time and spatial-frequency analysis, commonly employed in imaging applications such as filtering, compression, and reconstruction, and the operation has been demonstrated in previous works\autocite{zhao2021implementation,zhao2023energy}. 
Despite showing speed and energy advantages, the non-idealities of memristor devices, such as stuck-at-faults and programming variations, still significantly degrade the fidelity of processed images, and since the matrix is pre-defined, the performance cannot be improved by neural network training. 
Here, we demonstrate that our memristor can tackle this limitation.
Our hardware demonstration is illustrated in \autoref{Fig:app1}a, where the 2D images are in smaller tiles to align with the size of each memristor crossbar. 
For an $N \times N$ input $\mathbf{x}$, the 2D-DFT is computed as two successive matrix multiplications using 1D-DFTs. Mathematically, this is expressed as: $Y = [\mathbf{W_{1D-DFT}}(\mathbf{W_{1D-DFT}}\mathbf{x})^T]^T$.
In our implementation, this process was decomposed into three in-memory steps: performing an in-memory 1D-DFT, transposing the result, and then applying another in-memory 1D-DFT on the same crossbar.
Both real and imaginary components of the 32-point DFT matrix are programmed to to crossbar array for in-memory image processing experiments.

To evaluate the practical programming performance of the proposed method against the traditional method, we compared both approaches with different configurations: traditional non-adaptive, differential pair matrix representation (with and without compensation layers) and our proposed adaptive, fault-free, and differential pair-free method (with and without compensation layer).
The required memristor device count for both methods is 4,096 without compensation array, while 8,192 with compensation array. 
The experimentally represented matrices were rescaled for DFT matrix values from the conductance values, and the program error is evaluated by the error in the represented matrix, as shown in \autoref{Fig:app1}b.
The results show that the traditional method exhibits frequent extreme errors even with a compensating layer, which partially reduces variations but cannot eliminate fault-induced outliers.
In contrast, our method suppresses faults through mathematical optimization in matrix representation by adaptively aligning with faulty-device values, resulting in lower error magnitudes and a more uniform distribution. 
Remarkably, our design without compensation reduces extreme errors to negligible levels (no visible outliers in \autoref{Fig:app1}b). Adding a compensating layer to it further narrows the error distribution, but its contribution is incremental to its intrinsic fault tolerance.

To visualize the effectiveness of our methods, we reconstructed the images using an ideal two-dimensional inverse Discrete Fourier Transform (2D-IDFT) from the 2D frequency domain (\autoref{Fig:app1}c) by the experimental DFT operations in crossbars. 
The comparison reveals a substantial improvement in image quality achieved, as seen in the sharper details and reduced artifacts compared to the traditional method. 
These improvements are further quantified in \autoref{Fig:app1}d, where the results demonstrate that our method improves the signal-to-noise ratio (SNR) by 28.9 dB ($776\times$ signal-to-noise ratio) when using the same hardware resources with the traditional method.

Furthermore, to investigate the resource efficiency of our design, we also conducted experimental programming of both real and imaginary components of 32-point DFT matrices with k=17, implemented both without compensation array (utilizing 2176 devices) and with compensation array (utilizing 4352 devices). Experimental results conclusively demonstrate that our methodology enables a 47\% reduction in device count while simultaneously delivering a 15.9 dB ($39\times$ signal-to-noise ratio) enhancement in SNR, thus confirming its superior efficiency-performance characteristics.

\section*{In-memory Baseband Processing for Wireless Communication.} 
\begin{figure}[H]
\centering
\includegraphics[width=\textwidth]{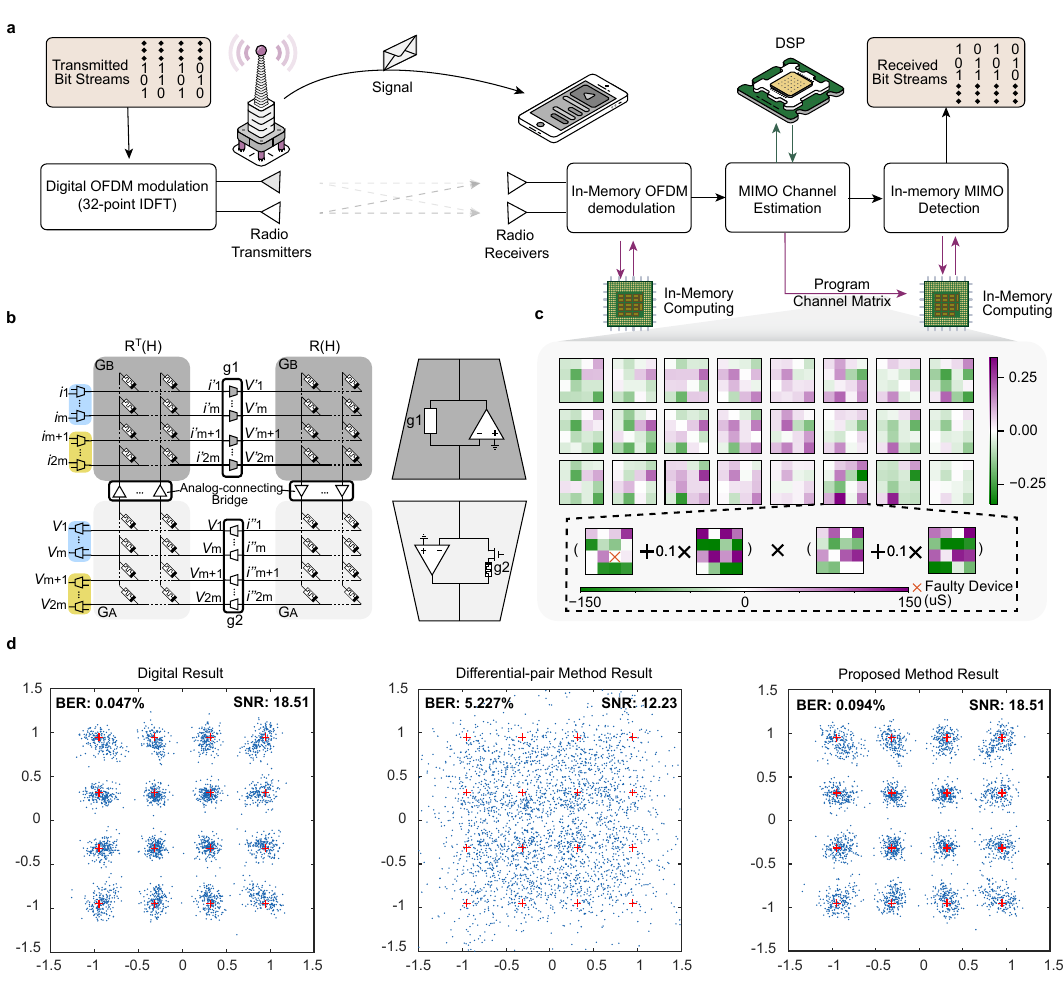}
\caption{
{\textbf{In-memory baseband processing for wireless communication.} }
{\textbf{a},} Schematic of the in-memory baseband processing architecture, supporting simultaneous multi-antenna transmission (MIMO) and multi-frequency transmission (OFDM) to enhance spectral efficiency and system capacity. 
{\textbf{b},} Implementation of a high-precision MIMO detector based on the proposed method. 
{\textbf{c},}  MIMO channel matrix programmed using the proposed method. Each $2\times2$ complex matrix is transformed into a $4\times4$ real-valued matrix and represented as the product of two $4\times4$ submatrices. 
Each submatrix is equipped with a compensation layer to mitigate programming variation. 
{\textbf{d}, Constellation diagrams for the benchmark digital receiver (SNR: 18.51 dB, BER: 0.047\%), the in-memory receiver with direct mapping (SNR: 12.23 dB, BER: 5.227\%), and the in-memory receiver using the proposed method (SNR: 18.51 dB, BER: 0.094\%), demonstrating the effectiveness of our design in maintaining signal quality.}
}
\label{fig:baseband}
\end{figure}

Baseband processing for efficient sixth-generation (6G) wireless communication is another domain that requires efficient and accurate computing with non-trainable matrix parameters.
Our recent work has demonstrated that with the in-memory computing capacity enabled by memristors, the communication can achieve speeds of $160.8$ Gb/s ($671\times$ faster than digital state-of-the-art) with $0.22 pJ$/bit energy efficiency ($329\times$ improvement)\autocite{zeng2023realizing}, but at an expense of reduced precision by imperfect matrix representation caused by device non-idealities, which is the key hurdle that prevents from practical deployment.

To address this gap, we implement a 6G signal transmission prototype between transmitters and receivers (see \autoref{fig:baseband}a), where our proposed adaptive matrix representation enables precision-guaranteed acceleration of two key techniques in receivers: orthogonal frequency division multiplexing (OFDM) and multi-input multi-output (MIMO).

Orthogonal frequency division multiplexing (OFDM) is a key technique for mitigating frequency-selective fading in wireless communication by partitioning bandwidth into multiple narrow, orthogonal subchannels. 
In our 6G prototype, we divided the available bandwidth into 32 orthogonal subchannels, allocating 24 high-quality channels for data transmission while leaving 8 lower-quality channels unused (including one direct current frequency channel affected by direct current leakage and edge channels impacted by interference and filter distortion). 
After receiving the raw signals, the Fourier transform is performed to decode the information carried in a particular channel (frequency). The DFT also enables the decoding of arbitrarily selected frequencies, a capability that the FFT cannot satisfy. 
For processing the complex-valued DFT computations, we mapped real and imaginary components onto separate $32 \times 32$ crossbars using our adaptive matrix representation method. 
Each submatrix is accompanied by a compensation array to reduce the impact of device variations, thereby reducing computational noise during in-memory OFDM processing.

MIMO technology further increases throughput by transmitting multiple data streams simultaneously across multiple antennas. However, this simultaneous transmission creates mutual interference between data streams at the receiver. 
Therefore, to recover the intended data stream on each OFDM subchannel via a specific antenna pair, a MIMO detector needs to recover the transmitted vector using an estimated channel matrix $\hat{\mathbf{H}}$, as detailed in Methods. 
This step is computationally expensive, with a computational complexity of $O(N^3)$  stemming from matrix inversion using Gaussian elimination, where $N$ represents the number of antennas, thereby becoming one of the most prominent bottlenecks when they are implemented in energy-efficient demanding scenarios. 
Here, we implement the matrix inversion operation in the MIMO detector with memristor crossbar using analog-connected continuous-time iteration to accelerate the operation, and redesign the architecture (provided in \autoref{fig:baseband}b and detailed in Methods ) by our adaptive matrix representation method to mitigate the impact of device imperfections. 
We use a $2 \times 2$ MIMO system (two transmit antennas and two receive antennas) as a proof-of-concept, and the estimated channel matrix was mapped and programmed to memristor crossbars using our proposed matrix representation method. 
Specifically, the complex-valued matrix $\mathbf{H}$ is mapped as:
$
\begin{bmatrix}
\Re{(\mathbf{H})} & -\Im{(\mathbf{H})} \\
\Im{(\mathbf{H})} & \Re{(\mathbf{H})}
\end{bmatrix}
$\autocite{zhao2023energy}. 
The experimentally memristor programming result is shown in \autoref{fig:baseband}c, with each $4 \times 4$ transformed matrix divided into the product of two $4 \times 4$ submatrices, with each submatrix accompanied by one compensation layer, using a total of 1,536 memristor devices. 
The result shows that the final channel matrix represented by our experimentally programmed memristors has a cosine similarity of over 99.9\% with the corresponding ideal matrix, confirming the effectiveness of our proposed approach.

After redesigning both the OFDM and the MIMO detector with our proposed adaptive matrix representation method, we are able to achieve significantly enhanced baseband processing quality. 
To showcase the improvement compared to previous work, we present the constellation diagrams of the receiving data processed by different baseband signal processors, including the digital receiver as the baseline, the in-memory receiver using traditional non-adaptive, differential-pair matrix representation, and the in-memory receiver using the proposed adaptive, fault-free matrix representation (\autoref{fig:baseband}d). 
The quality of baseband receivers is indicated by the dispersion of the received data from the reference points on the diagram - the more concentrated, the better the quality. 
The data of the digital benchmark and the proposed scheme are more concentrated around the reference points, nevertheless, with a narrow dispersion, which is mainly caused by the signal noise. 
In contrast, the differential-pair in-memory receiver demonstrates a wider dispersion, primarily caused by significant hardware noise introduced during in-memory signal processing. 
This heaviness limits the practical adoption of this technology, although our previous work has demonstrated significant speed and energy benefits. 
To quantitatively assess performance, we measured the Bit Error Rate (BER, lower is better) and Signal-to-Noise Ratio (SNR, larger is better) for each method. Our in-memory receiver achieved an SNR of 18.51 dB, matching the digital receiver and outperforming the differential-pair in-memory receiver by 6.28 dB. More importantly, the proposed receiver exhibited a BER of 0.094\%, representing a 56-fold reduction compared to the differential-pair method (5.227\%) and nearing the digital receiver's 0.047\% BER. 
These results further validate the feasibility of the proposed scheme for accelerating practical, precision-sensitive tasks, such as in-memory baseband signal processing.

\section*{Scaled architecture and performance benchmark on real-world applications}
\label{sec:benchmark}
\begin{table}[H]
\caption{Performance comparison between differential-pair matrix representation and the proposed design}
\label{tab:benchmark}
\begin{tabular*}{\columnwidth}{@{\extracolsep\fill}lccc|ccc}
\Xhline{1.2pt} 
\toprule%
& \multicolumn{3}{@{}c@{}}{\makecell{Differential-pair matrix representation \\\textbf{(Vulnerable to stuck-at-fault issues)}}} & \multicolumn{3}{@{}c@{}}{\makecell{This work \\\textbf{(Immune to stuck-at-fault issues)}}} \\\cmidrule{2-4 
  }\cmidrule{5-7}%

  &device number & Area & Energy  &device number & Area & Energy\\
\Xhline{1.2pt}
Generic Matrix $\mathbf{M}_{m\times n}$   & $2mn$ & - & -  & $\min{\{m,n\}} \times (m+n)$ & - & -\\ \hline
DFT 256   & \makecell{262.1K \\($\times$ 1)} &  \makecell{0.493$mm^2$\\($\times$ 1)} & \makecell{50.52$nJ$\\($\times$ 1)}  &  \makecell{131.1K \\(\textbf{$\times$ 0.50})}  & \makecell{0.248$mm^2$\\(\textbf{$\times$ 0.50)}} & \makecell{25.96$nJ$\\(\textbf{$\times$ 0.51)}}\\\hline

\makecell{Transition matrix\\(Harvard 500)}  & \makecell{500K \\($\times$ 1)} & \makecell{0.498$mm^2$\\($\times$ 1)} & \makecell{75.57$nJ$\\($\times$ 1)} & \makecell{170K \\(\textbf{$\times$ 0.34})} & \makecell{0.255$mm^2$\\(\textbf{$\times$ 0.51)}} &
\makecell{29.73$nJ$\\(\textbf{$\times$ 0.39)}}\\\hline

VGG-16   & \makecell{263.9M \\ ($\times$ 1)} & \makecell{548.2$mm^2$\\($\times$ 1)} & \makecell{54.3$\mu J$\\($\times$ 1)} &   \makecell{147.8M \\(\textbf{$\times$ 0.56})} & \makecell{281.0$mm^2$\\(\textbf{$\times$ 0.51)}} &
\makecell{30.48$\mu $\\(\textbf{$\times$ 0.56)}}\\\hline
   
ResNet-18 & \makecell{23M  \\($\times$ 1)}     & \makecell{64.12$mm^2$\\($\times$ 1)} &
\makecell{5.12$\mu J$\\($\times$ 1)}  &   \makecell{12.4M\\(\textbf{$\times$ 0.58})}  & \makecell{32.94$mm^2$\\(\textbf{$\times$ 0.51)}} & \makecell{2.832$\mu J$\\(\textbf{$\times$ 0.55)}}\\
\Xhline{1.2pt}
\end{tabular*}
\footnotetext[1]{Example for a first table footnote.}
\footnotetext[2]{Example for a second table footnote.}
\end{table}
Many practical applications, such as those used in large language models (LLMs), require computing with much larger matrices than what a single-crossbar can accommodate. 
This size limitation of the single crossbar comes from factors like the need for flexibility to fit different tasks while maintaining reasonable utilization rates (same factor also considered in digital accelerators\autocite{google_tpu_architecture}), the parasitic IR drop along connection wires\autocite{xiao2021analysis}, the maximum current that can be supported by the peripheral circuitry\autocite{wan2022compute}, and so on.
To handle these large matrices, a common approach is to partition them into smaller, independent submatrices (tiling).
Our method is fully compatible with this tiling architecture and offers significant advantages, particularly when deploying trainable matrices such as those in neural networks that utilize techniques like in-situ training to mitigate defects such as stuck-at-faults \autocite{li2018efficient}. 
These benefits supplement the advantages previously shown for computations with non-trainable matrices.
Specifically, two key benefits emerge from combining our method with tiling:
First, because tiling allows each submatrix to be processed independently, the adaptive compensation of defects can be performed in parallel across all submatrices. 
This significantly reduces the overall time required for compensating hardware defects. 
Second, it provides a solution for handling runtime device faults during edge deployment, even in resource-constrained environments. When faults occur, only the affected submatrices need redeployment using our method, eliminating the need to reconfigure the entire system.

To evaluate the practical advantages of our approach beyond precision improvements, we analyze several key hardware implementation metrics that directly impact deployment feasibility and efficiency across various applications.
Key performance metrics, including required memristor device count, chip area, and energy consumption, are summarized in \autoref{tab:benchmark}. 
The evaluations cover a generic $m\times n$ matrix and representative workloads, including the 256-point Discrete Fourier Transform (DFT) used in image reconstruction tasks\autocite{zhao2021implementation,zhao2023energy}, the Harvard 500 transition matrix for PageRank computations\autocite{pedretti2021redundancy}, and two widely used convolutional neural networks—VGG-16 and ResNet-18—employed for classification tasks\autocite{simonyan2014very,he2016deep,yao2020fully}. 
Details on how these benchmark results were obtained are provided in  Supplementary Note 3.2. 
Compared to the previous differential pair matrix representation, our proposed method significantly reduces the required memristor device count.
This reduction resulted from eliminating differential pairs and compressing information at the intermediate node between the two compensation arrays. 
For specific applications, this leads to a device count reduction to 57.8\% to 34.0\% of that of the baseline.
More generally, for any large matrix, this method generally reduces the device count to 50\% of the baseline (as proven in  Supplementary Note 3.3), but the number can be higher if imperfect tiling, or lower for special matrics properties.
The presence of exploitable structures in certain matrices, like low-rankness or sparsity, presents opportunities for further optimization, motivating future research into specialized algorithms. 
Primarily driven by this reduction in device count, along with associated circuit-level co-optimizations enabled by our method, we also achieve up to 200\% improvement in area efficiency and up to 256\% in energy efficiency.
While the computation latency slightly increases because signals pass through two arrays, the overall impact is minimal. 
This is because the connection between the arrays avoids the expensive analog-to-digital conversion step, which constitutes the most significant overhead in traditional designs.
Our analysis shows that the latency is $1.16\times$ that of the traditional method, primarily due to signal delays through several amplifiers in the analog connection circuit.
This minor latency increase is generally considered an acceptable trade-off for the considerable gains in energy efficiency and defect tolerance, particularly crucial in energy-sensitive edge computing scenarios.

\section*{Discussion}
In summary, we have introduced and validated a novel adaptive indirect matrix representation method that effectively mitigates the negative impact of stuck-at-faults and device variations inherent in analog in-memory computing hardware. 
By decomposing a target matrix into the product of two adjustable sub-matrices, our approach enables robust computation even on highly imperfect physical substrates.
Our analysis and experiments on integrated memristor-based system revealed that near-perfect matrix representation (a greater than 99.999\% cosine similarity) for a Discrete Fourier Transform matrix can be achieved with a up to a 39\% device fault rate, and demonstrated a 56-fold reduction in bit-error-rate for a wireless communication prototype. 
Furthermore, this technique yielded over 196\% improvement in computational density and over 179\% improvement in energy efficiency compared to conventional non-adaptive, direct matrix representation. 
The successful application of this technique to different tasks showcased its practical viability. 
This work not only provides a crucial solution for current memristor-based systems but also offers a generalizable method applicable to a wide range of emerging memory technologies and even non-electrical computing paradigms, thereby paving the way for reliable and efficient analog computing in an era of ever-increasing computational demands.





\section*{Method}

\setcounter{section}{0}
\renewcommand{\thesection}{Method \arabic{section}}
\subsection*{Memristor integration}
We use an integrated Ta/\ce{TaOx}/Pt memristor hardware to demonstrate our idea, but the method is also applicable to other technologies.
The memristor devices are integrated on top of CMOS selectors and peripheral circuits manufactured in a commercial foundry using a 180 nm process. 
The integration is performed in house with back end processes, which begins by removing the native oxide on the surface metal through reactive ion etching (RIE) and a dip in buffered oxide etch (BOE). 
Next, a 2 nm chromium (Cr) adhesion layer and a 10 nm platinum layer are sputtered and patterned via e-beam lithography to form the bottom electrode. A 2 nm layer of tantalum oxide (\ce{TaOx}) is then reactively sputtered to serve as the switching layer,  followed by the sputtering of a 10 nm tantalum (Ta) layer for the top electrode. The device stack is finalized with a 10 nm platinum (Pt) layer, providing passivation and improved electrical conduction. 

\subsection*{Iterative write-and-verify programming method}
In this work, memristor devices are programmed to target conductance values (0--150~$\mu$S) using an iterative write-and-verify method with a $\pm$15~$\mu$S tolerance range. At each iteration, a 0.2~V READ pulse first measures the conductance. If the value lies below (target $-$15~$\mu$S) or above (target $+$15~$\mu$S) the tolerance range, a SET (1~$\mu$s pulse) or RESET (1~$\mu$s pulse) is applied, respectively; devices within the range remain unperturbed. During the programming process, both the programming voltage and gate voltage are gradually increased, as detailed in Table S1.

\subsection*{Hardware-aware matrix decomposing via gradient-based optimization}
\label{method:dataset free training}
The matrix decomposition process is formalized through an iterative gradient-based optimization framework. 
As illustrated in Figure S1, the identity matrix $I \in \mathbb{R}^{m \times m}$ serves as the fixed input to the system, undergoing sequential transformation through parameter matrices $M_A \in \mathbb{R}^{m \times k}$ and $M_B \in \mathbb{R}^{k \times n}$. Specifically, the forward computation is expressed as:
\begin{equation}
\hat{M} = (I \cdot M_A) \cdot M_B = M_A M_B
\end{equation}

This reconstructed matrix $\hat{M}$ is then compared with the target matrix $M \in \mathbb{R}^{m \times n}$ through the previously defined cosine similarity loss:
\begin{equation}
\mathcal{L} = 1 - \frac{\langle \text{vec}(M_A M_B), \text{vec}(M) \rangle}{\|\text{vec}(M_A M_B)\| \cdot \|\text{vec}(M)\|}
\end{equation}

The Adam optimizer with learning rate $\eta = 1 \times 10^{-4}$ drives the parameter update process. At each iteration, gradients $\nabla_{M_A}\mathcal{L}$ and $\nabla_{M_B}\mathcal{L}$ are computed via automatic differentiation, followed by momentum-accelerated weight updates:
\begin{equation}
\begin{aligned}
M_A^{(t+1)} &\leftarrow M_A^{(t)} - \eta \cdot \hat{g}_A^{(t)}, \
M_B^{(t+1)} &\leftarrow M_B^{(t)} - \eta \cdot \hat{g}_B^{(t)}.
\end{aligned}
\end{equation}
where $\hat{g}_A^{(t)}$ and $\hat{g}_B^{(t)}$ represent gradient estimates. 

Two hardware-aware constraints are applied during updates. First, row-wise sign consistency is enforced through a polarity projection mechanism. For each row vector $M_A[i,:]$, we maintain a predetermined scalar sign pattern $s_i \in \{-1,+1\}$ using element-wise rectification:
\begin{equation}
M_A[i,:] \leftarrow s_i \cdot \max\left(0,\, s_i \cdot M_A[i,:]\right)
\end{equation}
This dual multiplication structure ensures elements maintain their designated polarity. 
Second, a device reliability constraint preserves predefined faulty positions through a masking operation:
\begin{equation}
M_A \leftarrow M_A \odot \Omega + M_A^{(0)} \odot (1-\Omega)
\end{equation}
where $\Omega \in \{0,1\}^{m \times k}$ is a binary mask preserving faulty devices, $M_A^{(0)}$ contains the initial values of faulty devices, and $\odot$ denotes the Hadamard product (element-wise multiplication). The optimization executes 5,000 epochs to enforce $M_AM_B=M$

\subsection*{Modeling memristor chip with stuck-at-fault devices}
We formalize the model for memristor chips with stuck-at-fault devices through mathematical representations of device failures and conductance constraints. Let the crossbar conductance matrix be defined as $\mathbf{G} \in \mathbb{R}^{m \times n}$, where each element $G_{ij}$ corresponds to the conductance of the device at row $i$ and column $j$. Faulty devices are modeled as follows:
\begin{equation}
G_{ij} = 
\begin{cases}
0, & \forall (i,j) \in \mathcal{F}_{\mathrm{OFF}} \quad \text{(Stuck-at-OFF)} \\
G_{\mathrm{max}}, & \forall (i,j) \in \mathcal{F}_{\mathrm{ON}} \quad \text{(Stuck-at-ON)}
\end{cases}
\end{equation}
where $G_{\mathrm{max}} = 150\ \mu\mathrm{S}$ denotes the maximum programmable conductance because of the series connected transistor selectors, with $\mathcal{F}_{\mathrm{OFF}}$ and $\mathcal{F}_{\mathrm{ON}}$ representing the sets of devices permanently locked in zero-conductance and maximum-conductance states, respectively. 
The fault sets $\mathcal{F}_{\mathrm{OFF}}$ and $\mathcal{F}_{\mathrm{ON}}$ are generated through sequential probabilistic selection: first, $\lfloor r_{\mathrm{OFF}} \cdot m \cdot n \rfloor$ distinct coordinates are randomly chosen from the $m \times n$ grid to form stuck-at-OFF failures; subsequently, $\lfloor r_{\mathrm{ON}} \cdot m \cdot n \rfloor$ distinct coordinates are selected from the remaining $m n - \lfloor r_{\mathrm{OFF}} \cdot m \cdot n \rfloor$ non-faulty devices to establish stuck-at-ON failures.

\subsection*{Precision evaluation by cosine similarity}
We quantify the deviation between the ideal matrix $\mathbf{M}$ and its nonideal-affected result $\mathbf{M}'$ using cosine similarity, defined as:
\begin{equation}
    \mathrm{Similarity}(\mathbf{M}, \mathbf{M}') = \frac{\langle \mathrm{vec}(\mathbf{M}), \mathrm{vec}(\mathbf{M}') \rangle}{\|\mathrm{vec}(\mathbf{M})\|_2 \cdot \|\mathrm{vec}(\mathbf{M}')\|_2}
\end{equation}
where $\mathrm{vec}(\cdot)$ denotes column-wise vectorization, $\langle \cdot, \cdot \rangle$ represents the inner product, and $\|\cdot\|_2$ is the Euclidean norm. The similarity metric naturally bounds values in $[-1, 1]$, with $1$ indicating identical matrices, $-1$ corresponding to diametrically opposed patterns, and values approaching $1$ reflecting stronger similarity. This formulation isolates the fault modeling mechanics from specific computational tasks, establishing a generalizable evaluation standard. 

\subsection*{Hardware implementation for the fault-tolerant and precision-controlled matrix representation}
Our hardware architecture design, illustrated in \autoref{Fig:FFMR_arbitrary}(a), implements the formulation $\mathbf{M_{m\times n}}=\sum_i^N{{k_A}_i \mathbf{C_A}_{i,m\times k}}\times \sum_i^N{{k_B}_i \mathbf{C_B}_{i,k\times n}}$, which combines adaptive matrix decomposition with compensation techniques to achieve fault-tolerant and precision-controlled programming. In our design, voltage input signals $\mathbf{x}$ first enter the submatrix $\mathbf{M_A}_{m\times k}$ layers, where they are converted into currents through memristor compensation layers. Each compensation layer ($\mathbf{C_A}_{i,m\times k}$) processes these currents via operational amplifiers with specific feedback resistors ${R_{fA}}_i$, generating scaled currents according to the ratio ${k_A}_i = {R_{fA}}_i / R$. Currents from multiple layers are accumulated to form the representation of $\mathbf{x}\mathbf{M_A} = \mathbf{x}\sum_i^N{{k_A}_i \mathbf{C_A}_i}$. These accumulated results are then transmitted through an analog connecting bridge to submatrix $\mathbf{M_B}$. In this second stage, a similar process with feedback resistors ${R_{fB}}_i$ implements the operation $(\mathbf{xM_A})\mathbf{M_B} = (\mathbf{xM_A})\sum_i^N{{k_B}_i \mathbf{C_B}_i}$, before the final output reaches the analog-to-digital converter.

\subsection*{Fault-tolerant and precision-controlled matrix representation algorithm}
\label{method:SAF_arbitrary_precision_programming}

By combine the adaptive matrix decomposition representation that handle the device defect and the compensation arrays that reduce the impact of inaccurate device programming, we can achieve the fault-tolerant and precision-controlled matrix representation at the same time.
To determine the memristor conductance representing each matrices, the process consists of three steps (illustrated in Figure S4):

\begin{enumerate}
    \item \textbf{Initialization:} Positions of stuck-at-fault devices in the initial compensation layers ($\mathbf{C_A}_1$ and $\mathbf{C_B}_1$) are identified, and matrices ($\mathbf{M_A}_1$ and $\mathbf{M_B}_1$) are optimized using adaptive matrix decomposition to mitigate these stuck-at-fault devices. 
    The conductance values $\mathbf{G_A}_1$ and $\mathbf{G_B}_1$ for chip programming are determined by:
    \[
    \mathbf{G_A}_1 = \text{Sign}(\mathbf{M_A}_1) \cdot \max \left\{ 0, \left| \frac{\mathbf{M_A}_1}{{k_A}_1} \right| - \Delta G \right\},\quad \mathbf{G_B}_1 = \text{Sign}(\mathbf{M_B}_1) \cdot \max \left\{ 0, \left| \frac{\mathbf{M_B}_1}{{k_B}_1} \right| - \Delta G \right\}
    \]
    where $\Delta G$ is determined by empirical programming variation. 
    Specifically, in our implementation, the value is set to $3\sigma$ of the programming variation distribution observed in our experiment. 
    ${k_A}_1$ and ${k_B}_1$ are scaling factors constraining $\mathbf{G_A}_1$ and $\mathbf{G_B}_1$ within $0$ to $150 \, \mu S$.

    \item \textbf{Iteration:} After programming the initial conductance values onto the chip (obtaining $\mathbf{G_A}_1'$ and $\mathbf{G_B}'_1$), they are rescaled to refine matrices $\mathbf{M_A}_1'$ and $\mathbf{M_B}_1'$. For subsequent compensation layers (e.g., $\mathbf{C_A}_2$, $\mathbf{C_B}_2$), the process continues by ensuring that: 
    \begin{equation}
    \text{Target Matrix } \textbf{M} = \left(\sum_{j=1}^{i-1} \mathbf{M_A}_j' + \mathbf{M_A}_i \right) \times \left(\sum_{j=1}^{i-1} \mathbf{M_B}_j' + \mathbf{M_B}_i \right)
    \end{equation}
    remains accurate while also addressing stuck-at faults in the current layer. During iterations, two constraints are maintained: (a) sign consistency between consecutive layers ($\text{Sign}(\mathbf{M_A}_i) = \text{Sign}(\mathbf{M_A}_{i-1})$, $\text{Sign}(\mathbf{M_B}_i) = \text{Sign}(\mathbf{M_B}_{i-1})$), and (b) the ratio of maximum absolute values between consecutive layers remains below a specified threshold to minimize write variation influence:$\frac{\max|\mathbf{M_A}_i|}{\max|\mathbf{M_A}_{i-1}|} < \text{threshold ratio} \quad \text{and} \quad \frac{\max|\mathbf{M_B}_i|}{\max|\mathbf{M_B}_{i-1}|} < \text{threshold ratio}$.

    \item \textbf{Finalization:} For the final compensation layer, conductance values ($\mathbf{G_A}_N$, $\mathbf{G_B}_N$) are calculated without subtracting $\Delta G$ to achieve higher programming accuracy. The complete programming result is expressed as:
    \begin{equation}
    \text{Target Matrix } = \sum_i \mathbf{M_A}_i' \times \sum_i \mathbf{M_B}_i'
    \end{equation}
\end{enumerate}

\subsection*{Overview of communication system with in-memory baseband processing}

\label{method:mimoofdm_principles}
The proposed communication system with MIMO-OFDM with in-memory processing capability is designed to achieve high-throughput communication under a wideband rich-scattering wireless channel.
The high throughput is achieved by transmitting and receiving with multiple antennas simultaneously, i.e., the MIMO technology, while the symbol interference in the multi-path wideband channel is mitigated by the OFDM technology.
The wireless communication system is deployed in an environment where multiple propagation paths exist between the transmit antennas and receive antennas, through scattering, reflection, etc.. 
In the time domain, the transmitted signal will arrive at the receive antenna with several copies of different intensity and latency. In frequency domain, this results in a frequency-selective channel that has different gain at different frequencies. 
When wideband MIMO channel is decomposed into $N_c$ orthogonal sub-channels by the OFDM operations, each sub-channel is considered a narrowband channel that has a flat frequency response, the input-output response of any such sub-channel can be modeled as:
\begin{equation}
    \label{eq:channel_model}
    \mathbf{y}=\mathbf{Hx}+\mathbf{z},
\end{equation}
where, $\mathbf{x}\in\mathbb{C}^{N_t\times 1}$ and $\mathbf{y}\in\mathbb{C}^{N_r\times 1}$ denotes the symbol vectors transmitted and received, respectively, the sub-channel between $N_t$ transmitting antennas and $N_r$ receiving antennas can be characterized by a $N_r\times N_t$ matrix, denoted as $\mathbf{H}$.

\subsection*{OFDM algorithm in the communication system}
The time domain wireless channel is accurately modeled as convolution operations, however, the complexity at receiver to remove the convolution effect from signal is prohibitive. 
At the same time, circular convolution on time domain is equivalent to multiplication on frequency domain, which is much easier to handle at the receiver.
OFDM is such an algorithm that utilizes DFT/IDFT operations to transforms information between time and frequency domain. Information generated at the transmitter are considered to be on the frequency domain, then, IDFT operation is applied to transform the data into time domain. Tail of the generated time domain signal, called the cyclic prefix, is copied to pre-pend itself, such that the linear convolution channel is changed to a circular convolution one. This pre-pended signal is called an OFDM symbol. 
At the receiver, once the start of the OFDM symbol is correctly detected, the cyclic prefix is removed, and DFT is applied to transform the time domain information back to frequency domain, where the channel is estimated and removed.
The $N_c$-point IDFT equivalently slices the frequency domain wireless channel into $N_c$ sub-channels.
Both IDFT and DFT are matrix-vector multiplication operations.

\subsection*{Channel estimation algorithm in the communication system}
The wireless channel is considered stationary during a given time period, known as the channel coherence time, as a result, the channel needs to be estimated once in each coherence time at the receiver.
Since no information is transmitted during channel estimation, the estimation is considered an overhead and its frequency should be minimized to once per channel coherence time.
For each sub-channel, a pre-defined pilot symbol matrix $\mathbf{P}\in\mathbb{C}^{N_t\times N_t}$ is used for the estimation. 
Under the channel model \autoref{eq:channel_model}, denote the received symbol matrix as $\mathbf{R}\in\mathbb{C}^{N_r\times N_t}$, the channel matrix can be estimated using the maximum likelihood algorithm as: $\hat{\mathbf{H}}=\mathbf{R}\mathbf{P}^{\dagger}$, where $\mathbf{P}^{\dagger}$ denotes the inverse of $\mathbf{P}$. 
Channel estimation is a matrix-vector multiplication operation.

\subsection*{MIMO detection algorithm and hardware implementation with our proposed method}
When received symbol vector $\mathbf{y}$ and estimated channel $\mathbf{H}$ are known, the transmitted symbol vector $\mathbf{x}$ should be recovered, this is the MIMO detection process.
The MIMO detection algorithm adopted in this work is the popular MMSE algorithm\autocite{goldsmith2005wireless}, given by $\hat{\mathbf{x}} =(\hat{\mathbf{H}}^{\text{H}}\hat{\mathbf{H}} +(1/\text{SNR})\mathbf{I})^{-1}\hat{\mathbf{H}}^{\text{H}}\mathbf{y}$.
MIMO detection is a matrix-inverse-vector multiplication operation.

Consider the detection circuit with channel matrix
$\mathbf{H} \in \mathbb{C}^{Nr \times Nt} $.
The real mapped channel matrix $\mathcal{R}(\mathbf{H})^T \in \mathbb{R}^{2N_t \times 2N_r}$ is first mapped to the operational conductance range of memristor devices through scaling factor $\alpha$, yielding the programmable conductance matrix $\mathbf{G}=\mathbf{G_A}\mathbf{G_B} = \alpha \mathcal{R}(\mathbf{H})^T \in \mathbb{R}^{2N_t\times2N_r}$. 
This can be physically implemented using two memristor arrays connected in a cascaded multiplication configuration through an analog connecting bridge circuit that can choose the sign of input voltage by setting the sign register (see \autoref{Fig:MainIdea}e for analog connecting bridge circuit). The received signal vector $\mathcal{T}(y) \in \mathbb{R}^{2N_r\times1}$ is converted to input currents $i = \alpha \mathcal{T}(y)^T \in \mathbb{R}^{1\times2N_r}$, with $\alpha$ ensuring proper voltage swing within the circuit's operational limits. 

The left-side circuit configuration implements the matrix product $-\mathbf{G} = \mathbf{G_A} \times (-\mathbf{G_B})$ through voltage application at $\mathbf{G_A}$'s array terminals and sign inversion in the analog bridge circuitry. By applying Kirchhoff's current law, the summation of output currents from the left memristor chips ($-\mathbf{vG}$) and input currents ($\mathbf{i}$) produces: $\mathbf{i}' = -\mathbf{vG} + \mathbf{i}$. Subsequently, the transimpedance amplifiers (TIAs) generate output voltages: 
$\mathbf{v}' = (\mathbf{vG} - \mathbf{i})/g_1$
where $g_1$ denotes the TIA feedback conductance. The right part memristor arrays, implementing the transposed conductance matrix $\mathbf{G}^T = (\mathbf{G_A}\mathbf{G_B})^T = \mathbf{G_B}^T\mathbf{G_A}^T$, then performs the vector-matrix multiplication: $\mathbf{i}'' = \mathbf{v}'\mathbf{G}^T =(\mathbf{vG} - \mathbf{i}) \mathbf{G}^T/g_1$, Through the second TIA stage with feedback conductance $g_2$, we establish:
$\mathbf{v} = \mathbf{i}''/-g_2=(-\mathbf{vG} + \mathbf{i})\mathbf{G}^T/g_1g_2 \Rightarrow \mathbf{v}=\mathbf{iG}^T(\mathbf{GG}^T + g_1g_2\mathbf{I})^{-1}$
By setting the signal-to-noise ratio as $SNR = \alpha^2/(g_1g_2)$, the circuit implements the estimator:
$\mathcal{T}(\mathbf{\hat x})^T = \mathcal{T}(\mathbf{y})^T\mathcal{R}(\mathbf{H})[\mathcal{R}(\mathbf{H})^T\mathcal{R}(\mathbf{H}) + SNR^{-1}\mathbf{I}]^{-1} \Rightarrow \mathcal{T}(\mathbf{\hat x})=[\mathcal{R}(\mathbf{H})^T\mathcal{R}(\mathbf{H}) + SNR^{-1}\mathbf{I}]^{-1}\mathcal{R}(\mathbf{H})^T\mathcal{T}(\mathbf{y})$
\subsection*{Frame structure in the communication system}
For the MIMO-OFDM prototype, a proof-of-concept $N_t=N_r=2$ MIMO structure is adopted, 32-point DFT/IDFT is used in the OFDM, thus, $N_c=32$.
We adopt the frame structure illustrated in Figure S7, similar to the classic 802.11 WLAN protocol.
Several OFDM symbols and auxiliary signals are packed into one frame during transmission.
Each frame contains a preamble, a pilot, and a data payload. 
The preamble is a special signal with high auto-correlation, known by both the transmitter and the receiver. This signal is used for frame detection, time synchronization and frequency offset estimation\autocite{ieee2009ieee}.
The pilot signal consist of two OFDM symbol whose data is known a priori to the receiver, for each utilized sub-channel, data in the pilot signal represents a 2$\times$2 full rank matrix used to estimate the MIMO channel.
A number of OFDM symbols are concatenated together to form the data payload, the data payload contains the information to be transmitted to the receiver. 

\subsection*{Experimental setup for SDR signal reception}
Packets with random data symbols are generated in MATLAB at the PC that controls the SDR-Tx which then up-samples the baseband samples from 10M samples/s to 100M samples/s. After that, it
up-converts them to carrier frequency of 2.45GHz and emits the RF signal from its antenna. The signals then propagate over the air in
the lab environment. The SDR-Rx captures the signals, down-converts the RF samples to baseband, and re-samples the
baseband signals with a sample rate the same as the transmitter.
\renewcommand{\thesection}{\arabic{section}}
\section*{Acknowledgements}
The authors would like to thank Xunzhao Yin for the insightful discussions. 
This work was supported in part by the Research Grant Council of Hong Kong SAR (C7003-24Y, 27210321, C1009-22GF, T45-701/22-R, C5001-24Y), National Natural Science Foundation of China (62122005), ACCESS – an InnoHK center by ITC, and Croucher Foundation. 

\section*{Author contributions}
Z.X. and C. L. conceived the idea. Z.X. conducted experiments and performed simulations. J.L. conducted experiments on communication networks. S.H., Z.X., and Z.L. provided the mathematical proofs. C.L. supervised the project. Z.X., C.L., S.H., J.L., and S.W. wrote the manuscript. All authors contributed to the analysis of the results and provided comments on the manuscript. 
\section*{Competing interests}
The authors declare no competing interests. 
\section*{Data availability}
Source data for quality of matrix representation with different stuck-at OFF/ON rates and varying $k$ is available at \url{https://github.com/ZCXU0421/Fault_free_analog_computing.git}. Other data in this paper are also available with reasonable requests from the corresponding author.
\section*{Code availability}
The core code for the Fault-Free Matrix Representation and its integration with the compensation layer is publicly available at the GitHub repository: \url{https://github.com/ZCXU0421/Fault_free_analog_computing.git}.

\printbibliography[heading=subbibliography,title={References}]
\end{refsection}
\label{lastpage-main}
\clearpage
\fancyhf{}            
\fancyfoot[R]{\thepage/\pageref{lastpage-si}} 
\pagestyle{fancy}
\begin{refsection}


\printbibliography[heading=subbibliography,title={Supplementary References}]
\thispagestyle{fancy}

\end{refsection}
\label{lastpage-si}

\begin{appendices}






\end{appendices}



\end{document}